\documentclass[aps,prd,showpacs,eqsecnum,twocolumn,superscriptaddress,a4paper,nofootinbib]
{revtex4-1}
\usepackage{amsmath,amssymb,graphicx,color}
\usepackage{bm}
\usepackage{epsf}

\begin{document}

\title{Gravitational waves from very massive stars collapsing to a black hole}

\author{Haruki Uchida}
\affiliation{Yukawa Institute for Theoretical Physics, 
Kyoto University, Kyoto, 606-8502, Japan~} 

\author{Masaru Shibata} 
\affiliation{Max Planck Institute for
  Gravitational Physics (Albert Einstein Institute), Am Mühlenberg 1,
  Potsdam-Golm 14476, Germany}
\affiliation{Yukawa Institute for Theoretical
  Physics, Kyoto University, Kyoto, 606-8502, Japan~}


\author{Koh Takahashi}
\affiliation{Argelander-Institut f{\"u}̈r Astronomie, Universi{\"a}̈t Bonn, D-53121 Bonn, Germany}

\author{Takashi Yoshida}
\affiliation{Department of Astronomy, Graduate School of Science,
the University of Tokyo, Tokyo, 113-0033, Japan~} 

\date{\today}
\newcommand{\beq}{\begin{equation}}
\newcommand{\eeq}{\end{equation}}
\newcommand{\beqn}{\begin{eqnarray}}
\newcommand{\eeqn}{\end{eqnarray}}
\newcommand{\pa}{\partial}
\newcommand{\vp}{\varphi}
\newcommand{\varep}{\varepsilon}
\newcommand{\ep}{\epsilon}
\newcommand{\comp}{(M/R)_\infty}
\begin{abstract}

We compute gravitational waves emitted by the collapse of a rotating
very massive star (VMS) core leading directly to a black hole in
axisymmetric numerical-relativity simulations. The evolved rotating
VMS is derived by a stellar evolution calculation and its initial mass
and the final carbon-oxygen core mass are $320M_\odot$ and $\approx
150M_\odot$, respectively. We find that for the moderately rapidly
rotating cases, the peak strain amplitude and the corresponding
frequency of gravitational waves are $\sim 10^{-22}$ and $f \approx
300$--600\,Hz for an event at the distance of $D=50$~Mpc.  Such
gravitational waves will be detectable only for $D \alt 10$~Mpc by 
second generation detectors, advanced LIGO, advanced VIRGO, and KAGRA,
even if the designed sensitivity for these detectors is achieved. 
However, third-generation detectors will be able to detect such
gravitational waves for an event up to $D \sim 100$~Mpc.  The
detection of the gravitational-wave signal will provide a potential
opportunity for verifying the presence of VMSs with mass $\agt
300M_\odot$ and their pair-unstable collapse in the universe. 

\end{abstract}

\pacs{04.25.D-, 04.30.-w, 04.40.Dg}

\maketitle


\section{Introduction}

Gravitational collapse of a carbon-oxygen (CO) core induced by the
instability associated with the electron-positron pair creation (i.e.,
pair instability) is a possible fate of a very massive star (VMS) of
initial mass $\agt 140M_\odot$~\cite{FWH01,HW02,UM02,Takahashi16}. 
Broadly speaking, there are two possible fates for the
pair-instability collapse of the VMSs.  For the relatively low-mass
case, oxygen burning occurs explosively during the collapse, and then,
the thermal pressure resulting from the thermal energy released by the
thermal nuclear reaction of oxygen halts the collapse, leading to a
pair-instability supernova explosion (PISN).  On the other hand, for the
high-mass case, the collapse cannot be halted by the nuclear burning,
and hence, the final remnant is a black hole (BH) possibly surrounded by a
disk and an outflow~\cite{Uchida18}. However, PISN and/or BH formation 
have not been confirmed yet. 

In this article, we report a result of our new numerical-relativity
simulations for the collapse of a rotating VMS of its initial mass
$320M_\odot$ to a BH, paying attention to gravitational waves emitted
during the BH formation. We show that (i) the gravitational-wave
signal is characterized by a ringdown oscillation of the newly-formed
BH; (ii) for the moderately rapidly rotating cases, the energy emitted
by gravitational waves is typically $2 \times 10^{-7}M_{\rm CO}c^2$
where $M_{\rm CO}$ is the mass of the CO core just before the onset of
the collapse; (iii) the frequency for the peak amplitude of
gravitational waves is 300--600\,Hz for $M_{\rm CO} \approx
150M_\odot$ with the peak amplitude $\sim 10^{-22}$ for a hypothetical
distance to sources $D \sim 50$\,Mpc. Such gravitational waves can be
a target of third-generation gravitational-wave detectors such as
Einstein Telescope~\cite{ET}.

This paper is organized as follows. In Sec.~II, we summarize the setup
of our numerical-relativity simulation.  In Sec.~III, we present the
gravitational waveforms to show that these gravitational waves will be
a source for third-generation gravitational-wave detectors.
Section~IV is devoted to a summary and discussion.  Throughout this
paper, $c$ and $G$ denote the speed of light and gravitational
constant, respectively.


\section{Setup}


Following Ref.~\cite{Uchida18}, we employ an evolved rotating VMS
derived by a stellar evolution code of
Refs.~\cite{Takahashi16,Takahashi18} for preparing the initial
conditions for a numerical-relativity simulation.  The stellar
evolution calculation is performed for a metal-free star of its mass
$320M_\odot$ and of the rigid rotation with the angular velocity of
$2.1 \times 10^{-4}\,{\rm s}^{-1}$ at the zero-age main sequence
stage. The calculation is continued until the central temperature
reaches $\approx 10^{9.2}$\,K for which the star is composed of a CO
core (mainly of oxygen) and an envelope of helium and hydrogen (see
Fig.~2 of Ref.~\cite{Uchida18}).  At this stage, the total mass of the
entire star reduces to $\approx 290M_\odot$ because of the mass loss
during the stellar evolution~\cite{Yoon12}, and the mass of the CO
core is $\approx 150M_\odot$. The dimensionless spin parameter,
$cJ_{\rm CO}/(GM_{\rm CO}^2)$, of the CO core is $\approx 1.1$. 

In this paper, we explore the collapse of the CO core, reducing its
angular velocity uniformly to focus on the collapse of stellar cores
for which the dimensionless spin parameter is 0.1--0.9 (see
Table~\ref{tab1}).  For these cases, most of the matter in the CO core
eventually collapses into a rotating BH. In the following, we pay
particular attention to gravitational waves at the formation of the
BH, which reflect the early formation process of the BH.

\begin{table}[t]
\begin{center}
\caption{Quantities for the CO core of VMS employed in this paper.
  $M_{\rm CO}$: the mass.  $\beta$: ratio of rotational kinetic energy
  to gravitational potential energy.  $J_{\rm CO}$: angular momentum
  and $\chi:=cJ_{\rm CO}/(GM_{\rm CO}^2)$.  $R_{\rm CO}$: equatorial
  radius. \label{tab1} }
\begin{tabular}{ccccc}
\tableline\tableline
~Model~ & $M_{\rm CO}~(M_\odot)$ & $\beta$ & ~~~$\chi$~~~ & $R_{\rm CO}$\,(km) 
\\ \tableline
~M01~ & ~$150$~ & ~$2.7 \times 10^{-5}$~ & $0.11$ & $6 \times 10^5$ \\
~M03~ & ~$150$~ & ~$2.4 \times 10^{-4}$~ & $0.33$ & $6 \times 10^5$ \\
~M05~ & ~$150$~ & ~$6.6 \times 10^{-4}$~ & $0.55$ & $6 \times 10^5$ \\
~M07~ & ~$150$~ & ~$1.3 \times 10^{-3}$~ & $0.77$ & $6 \times 10^5$ \\
~M08~ & ~$150$~ & ~$1.7 \times 10^{-3}$~ & $0.88$ & $6 \times 10^5$ 
\\ \tableline 
\end{tabular}
\end{center}
\end{table}


Our method for the solution of Einstein's equation is the same as 
in Ref.~\cite{Uchida18}. We employ the original version of
Baumgarte-Shapiro-Shibata-Nakamura formulation with a puncture
gauge~\cite{BSSN}.  The gravitational field equations are solved in
the standard 4th-order finite differencing scheme.  The axial symmetry
is imposed using a 4th-order cartoon
method~\cite{cartoon,cartoon2,Uchida18}.  Gravitational waves are 
extracted from the outgoing-component of the complex Weyl scalar
$\Psi_{4}$, which is expanded by a spin-weighted spherical harmonics
of weight $-2$, $_{-2}Y_{lm}(\theta,\varphi)$, with $m=0$ in
axisymmetric spacetime (see, e.g.,~Ref.~\cite{yamamoto08}).  We focus
only on the quadrupole mode with $l=2$ (denoted by $\Psi_{20}$ in the
following) because it is the dominant mode (we checked that the 
amplitude of $l=3$ and $4$ modes is much smaller than $l=2$ mode). 

Formation of BH is determined by the presence of an apparent
horizon. The mass and dimensionless spin of the BH are
determined by measuring the area and circumferential radii of the
apparent horizon~(e.g., see Ref.~\cite{KST10} for our method). 

Following Ref.~\cite{Uchida18}, we use one of Timmes \& Swesty
equations of state which includes the contribution from radiation,
ions as ideal gas, electrons, positrons and corrections for Coulomb
effects. For electrons and positrons, the relativistic effect, the
effects of degeneration and electron-positron pair creation are taken
into account. We also take into account the nuclear burning,
photo-dissociation of iron and helium, and neutrino cooling.


The effect of the neutrino cooling is incorporated in the equation of
motion as
\beqn
\nabla_\alpha T^{\alpha\beta}=\rho u^{\beta} q_{\rm neu} \,f(\rho),
\eeqn
where $\nabla_\alpha$ is the covariant derivative with respect to the
spacetime metric, $T^{\alpha\beta}$ the energy-momentum tensor, $\rho$
the rest-mass density, $u^\alpha$ the four-velocity of the fluid,
$q_{\rm neu}$ the neutrino emission rate of Ref.~\cite{Itoh96}, and
$f(\rho)$ a function of $\rho$, respectively. Neutrinos freely escape
from the collapsing core for the case that the density and temperature
are not very high.  However, for $\rho \agt 10^{11}\,{\rm g/cm^3}$
with the temperature $T \agt 5$\,MeV, neutrinos with their cross
section $\agt 6 \times 10^{-43}(T/5\,{\rm MeV})^2\,{\rm cm}^2$ should
be trapped in the collapsing core of radius $\sim 10^7$\,cm.  To
approximately take into account this effect, we introduce a function,
$f(\rho)$, and set it as $\exp(-\rho/\rho_0)$ where $\rho_0$ is a
constant set to be $10^{11}\,{\rm g/cm^3}$ at the fiducial runs.  In
addition, we perform simulations for model M07 with
$\rho_0=10^{10}\,{\rm g/cm^3}$ and $\rho_0=0$ (i.e., $f=0$) to show
that the effect of the neutrino cooling plays a key role for
determining the spectrum of gravitational waves.

Numerical simulations are performed in cylindrical coordinates $(X,
Z)$, and a nonuniform grid is used for $X$ and $Z$.  Specifically, we
employ the following grid spacing (the same profile is chosen for $X$
and $Z$): for $X_i \leq X_{\rm in}$, $\varDelta X=\varDelta X_0=$const
and for $X_i > X_{\rm in}$, $\varDelta X_i=\eta \varDelta X_{i-1}$.
$\varDelta X_0$ is the grid spacing in an inner region and $\varDelta
X_i:= X_{i+1}- X_{i}$ with $X_i$ the location of $i$-th grid. $\eta$
determines the nonuniform degree of the grid spacing for which we
always choose 1.014.  As in Ref.~\cite{Uchida18}, for the early stage
of the collapse, we employ large values of $\varDelta X_0$ and $X_{\rm
  in}$ for which we assign the same values as before~\cite{Uchida18},
and then we perform a regriding for a better resolved simulation.  For
this later phase, we employ a grid resolution better than in our
previous study~\cite{Uchida18} as $\varDelta
X_0=0.0048$--$0.0096GM_{\rm CO}/c^2(\approx 1.1$--$2.2$)\,km. By
varying $\varDelta X_0$ for such a range, we confirmed that the
convergence of the gravitational waveform is well achieved.


\section{Numerical results}

\subsection{Black hole formation processes}

\begin{figure*}[t]
\epsfxsize=3.2in \leavevmode \epsffile{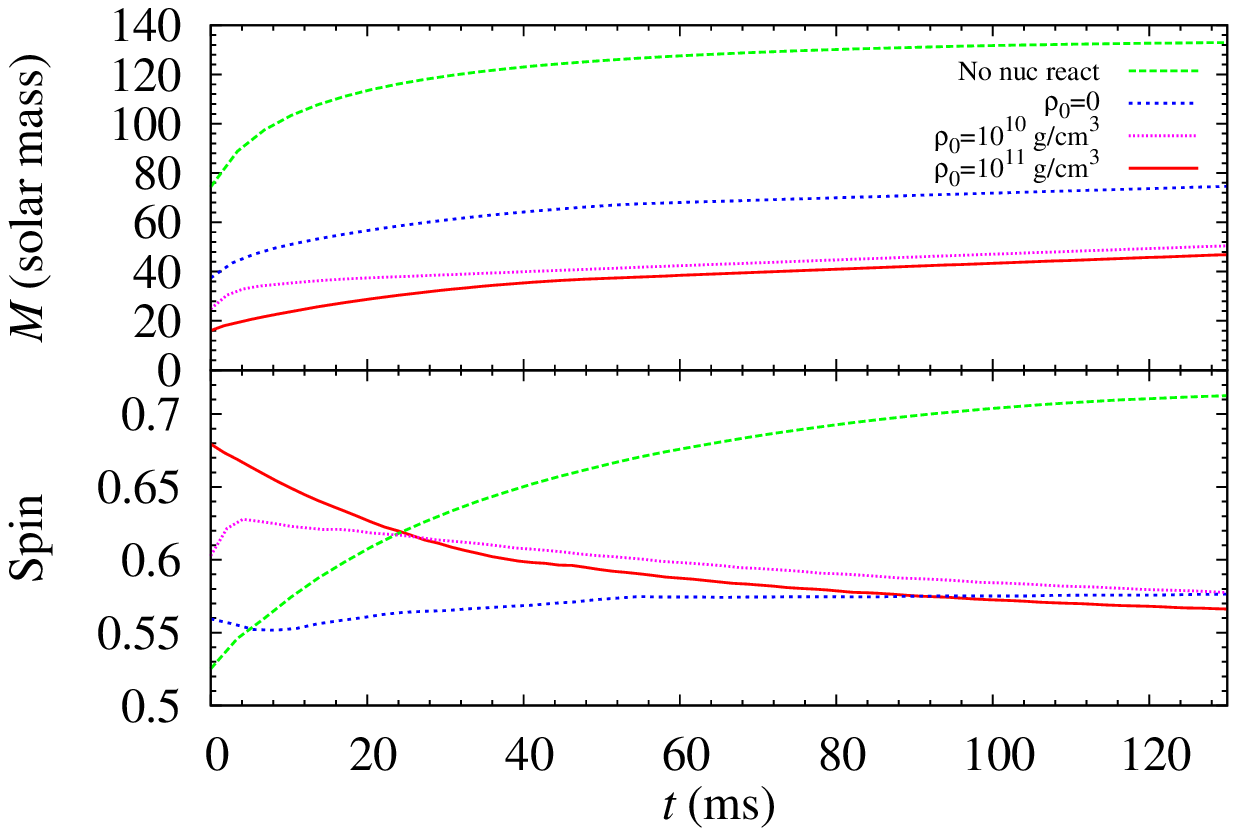}~~~~
\epsfxsize=3.2in \leavevmode \epsffile{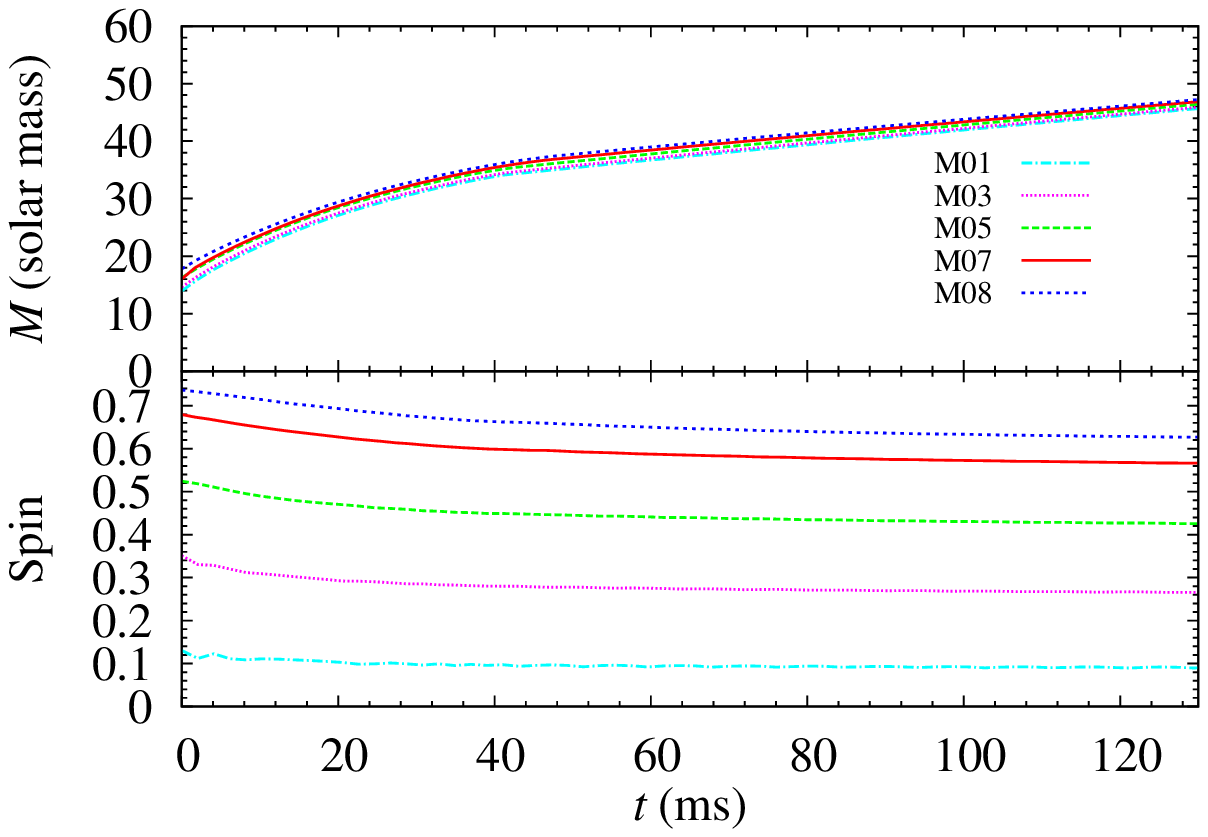}
\vspace{-2mm}
\caption{Left: Evolution of mass and dimensionless spin of BH
  formed after the collapse of VMSs for M07 models.  The
  dashed green, dotted blue, short-dotted magenta, and solid red
  curves show the results for no nuclear reaction and for the presence
  of nuclear reaction with $\rho_0=0$, $10^{10}$, and $10^{11}\,{\rm
    g/cm^3}$, respectively. We note that for the models in which the
  neutrino cooling is more efficient (for larger values of $\rho_0$),
  the initial BH mass is smaller.  Right: The same as the left
  panel but for models M01, M03, M05, M07, and M08 in the presence of
  the nuclear reaction and the neutrino cooling with
  $\rho_{0}=10^{11}\,{\rm g/cm^3}$.
\label{fig1}}
\end{figure*}

For all the models considered in this paper, a BH is formed in the CO
core collapse.  However, the formation and evolution processes of the
BH depend strongly on the effect of the nuclear reaction and neutrino
cooling. 

The left panel of Fig.~\ref{fig1} shows the evolution of the mass and
dimensionless spin for M07 models with no nuclear reaction and
$f(\rho)=0$ (no neutrino cooling), and with the nuclear reaction and
various values of $\rho_0$, $0$, $10^{10}$, and $10^{11}\,{\rm
  g/cm^3}$.  This shows that for no nuclear reaction, a substantial
fraction of the CO core collapses into a BH in the first $\sim
50$\,ms. By contrast, in the presence of the nuclear reaction (i.e.,
in the more realistic case), the initial mass of the BH is much
smaller than the total mass of the CO core. The reason for this is
that the photo-dissociation of iron and helium significantly reduces
the thermal pressure in the central region of the collapsing core, and
as a result, the collapse of the central region is significantly
accelerated, leading to a runaway collapse.

In addition to the nuclear reaction, the neutrino cooling plays an
important role in determining the initial BH mass, $M_i$. For no
neutrino cooling (i.e., $f(\rho)=0$ or $\rho_0=0$), $M_i\sim
40M_\odot$. On the other hand, if we take into account the neutrino
cooling ($\rho_0=10^{10}$ and $10^{11}\,{\rm g/cm^3}$), $M_i$ is
smaller as $\sim 20$--$30M_\odot$; i.e., for the larger value of
$\rho_0$, $M_i$ is smaller. The reason for this is that by the
neutrino cooling, the collapse is further accelerated in the central
region, leading to a smaller initial BH mass. 

By contract to the BH mass, the initial dimensionless spin is in a
fairly narrow range between 0.53 and 0.68, and thus, it does not
depend strongly on the effects of nuclear reaction and neutrino cooling
for the M07 models.  This property simply reflects the initial angular
momentum distribution of the CO core at the onset of the collapse. 

The right panel of Fig.~\ref{fig1} shows the evolution of the mass and
dimensionless spin for models M01, M03, M05, M07, and M08 with the
nuclear reaction and with the neutrino cooling for
$\rho_0=10^{11}\,{\rm g/cm^3}$.  This illustrates that the evolution
of the BH mass depends only weakly on the total angular momentum of
the collapsing star.  On the other hand, the BH spin naturally
reflects the angular momentum distribution of the CO cores.

\subsection{Gravitational waves}

\begin{figure*}[t]
\epsfxsize=3.3in
\leavevmode
\epsffile{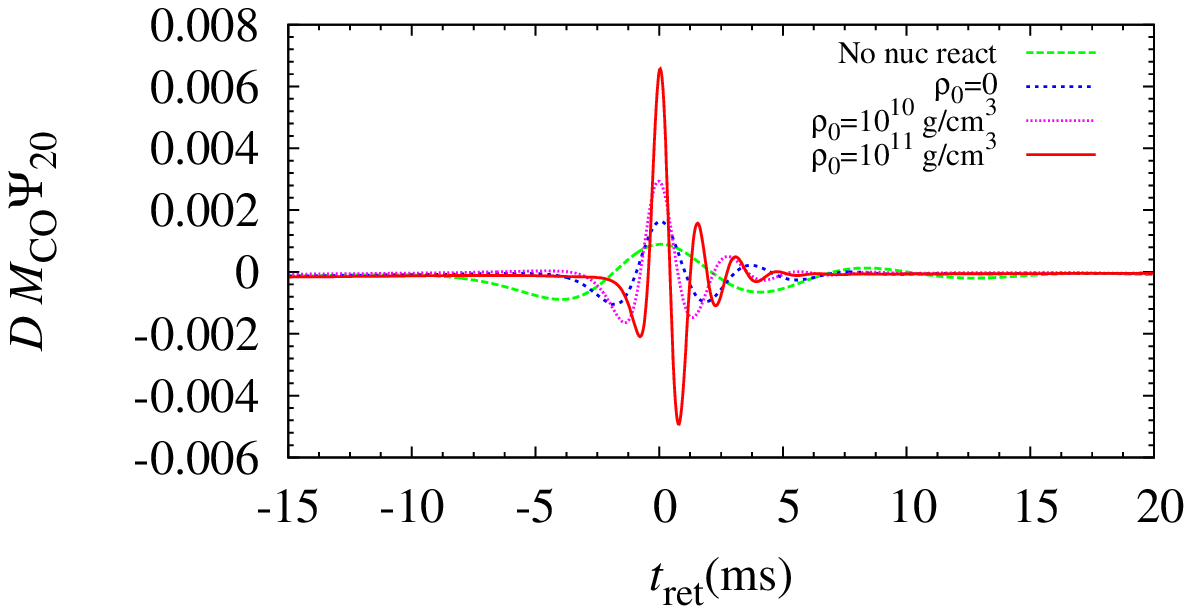}~~
\epsfxsize=3.3in
\leavevmode
\epsffile{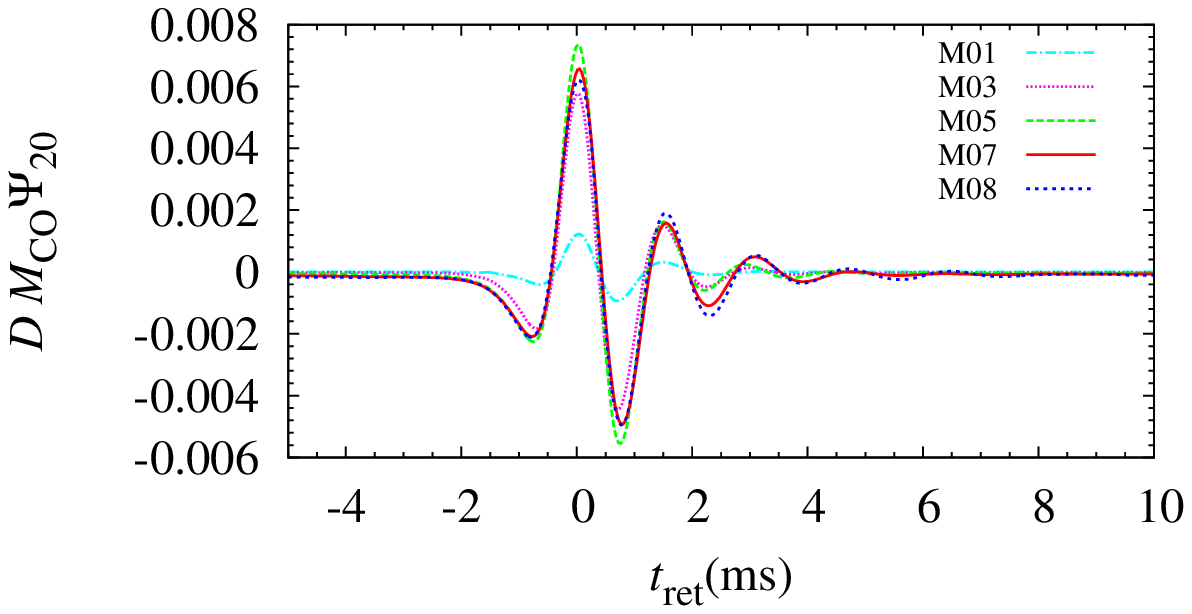}
\vspace{-3mm}
\caption{Left: Gravitational waveforms ($l=2$ axisymmetric mode of
  $\Psi_4$) as a function of retarded time for M07 models with no
  nuclear reaction (green dashed curve) and with nuclear reaction and
  neutrino cooling ($\rho_0=10^{11}\,{\rm g/cm^3}$: solid red curve,
  $\rho_0=10^{10}\,{\rm g/cm^3}$: short-dotted magenta curve, and
  $\rho_0=0$: dotted blue curve). Right: The same as the left panel
  but for models M01, M03, M05, M07, and M08 in the presence of the
  nuclear reaction and the neutrino cooling with
  $\rho_{0}=10^{11}\,{\rm g/cm^3}$.  For these figures, the time at
  which the maximum amplitude is reached is chosen as the origin of
  the time to align the waveforms.
\label{fig2}}
\end{figure*}

\begin{figure*}[th]
\epsfxsize=3.2in
\leavevmode
\epsffile{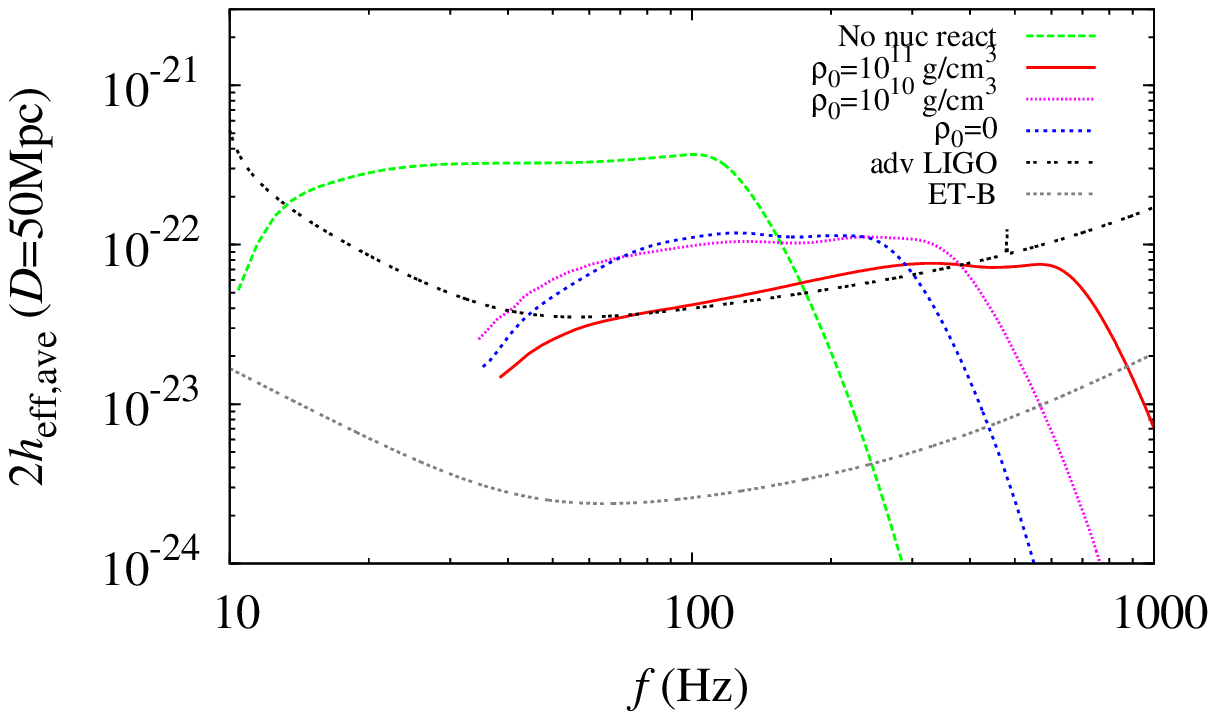}~~~~
\epsfxsize=3.2in
\leavevmode
\epsffile{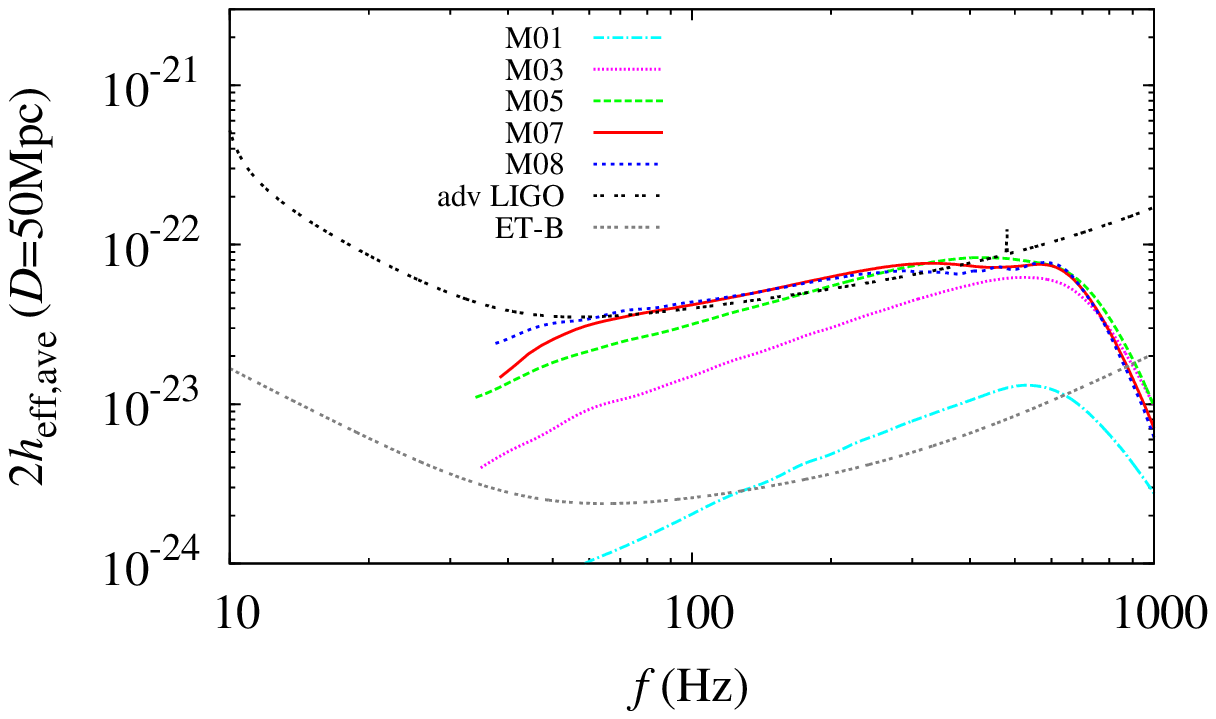}
\vspace{-3mm}
\caption{Left: Fourier spectrum of gravitational waves for a
  hypothetical distance to the source $D=50$\,Mpc for M07
  models. Here, we plot $2h_{\rm eff,ave}=2h_{\rm eff} \times
  2/3$. Each line type denotes the same model as in the left panel of
  Fig.~\ref{fig2}. The black dot-dot and grey dot-dot-dot curves show
  the design sensitivities of the advanced LIGO (the ``Zero Detuning
  High Power'' configuration~\cite{ligonoise}) and Einstein Telescope
  of the type B~\cite{etnoise}. Right: The same as the left panel but
  for models M01, M03, M05, M07, and M08. Each line type denotes the
  same model as in the right panel of Fig.~\ref{fig2}.
\label{fig3}}
\end{figure*}

Figure~\ref{fig2} displays gravitational waveforms ($\Psi_{20}$) as a
function of retarded time. Here, the time at which the maximum
amplitude is reached is chosen as the origin of the time.  We note
that $l=2, m=0$ mode is proportional to $\sin^2\theta$ where $\theta$
is the angle from the rotation axis.  Thus, the amplitude becomes
maximum for an observer located along the equatorial plane and it
vanishes if the observer is located along the rotation 
axis. Figure~\ref{fig2} plots the waveforms for $\theta=\pi/2$.

Figure~\ref{fig2} shows that gravitational waves are composed of a
short precursor and ringdown oscillation associated with the formed BH
for all the models considered in this paper.  The period of the
ringdown oscillation, $\lambda$, is varied for each model. We note
that for the $l=2, m=0$ mode, $\lambda$ should be $\approx 16GM_{\rm
  BH}/c^3 \approx 1.6(M_{\rm BH}/20M_\odot)$\,ms, where $M_{\rm BH}$
is the mass of the BH, according to the linear-perturbation analysis
for the BH quasi-normal modes~\cite{berti2009} with the dimensionless
spin, $\chi_{\rm BH}=0$--0.8.  For all the models with the
nuclear reaction, $\lambda$ agrees approximately with that predicted by
the BH perturbation theory if we take the initial mass of the BH as
$M_{\rm BH}$.  Thus, the gravitational waveforms reflect the early
formation process of the BH. For the model with no nuclear reaction
(dashed green curve in the left panel of Fig.~\ref{fig2}), $\lambda
\sim 16 \times (100GM_\odot/c^3) \sim 8$\,ms, reflecting the fact that
a substantial fraction of the CO core matter simultaneously collapses
into the BH. 

The right panel of Fig.~\ref{fig2} shows that the gravitational
waveforms depend weakly on the BH spin for $\chi \agt 0.5$. The main
reason for this is that the evolution process of the BH mass depends
only weakly on the spin.  For $\chi \alt 0.3$, the maximum amplitude
decreases steeply with the decrease of $\chi$. This reflects
the fact that for such small values of $\chi$, the degree of axial
symmetry decreases with $\chi$.  For higher BH spins, the damping time
for the ringdown oscillation is longer.  This also agrees with the
prediction by the BH perturbation theory~\cite{berti2009}.  For $\chi
\agt 0.5$, the peak amplitude is slightly higher for the lower spin.
This reflects the fact that for the higher spin, the collapse is
decelerated by a stronger centrifugal force, and the
gravitational-wave emission is slightly suppressed.


The total energy of gravitational waves emitted is $\Delta E \approx 7
\times 10^{-7}M_{\rm CO}c^2$ for the model with no nuclear reaction
and $\approx 2 \times 10^{-7}M_{\rm CO}c^2$ for the models with the
nuclear reaction and $\chi \agt 0.5$.  Thus, the emissivity is
suppressed in the presence of the nuclear reaction (specifically
photo-dissociation). The emissivity is much smaller than those in
binary BH mergers in which $\sim 10$\% of the total mass energy can be
radiated (e.g., Ref.~\cite{lovelace}). The emissivity for no nuclear
reaction is as large as that for the collapse of supermassive stars to
a BH~\cite{SSUU}.  For $\chi \alt 0.5$, $\Delta E$ steeply decreases
with $\chi$: $\Delta E/(M_{\rm CO}c^2) \approx 4 \times 10^{-9}$ and
$1 \times 10^{-7}$ for M01 and M03.

Figure~\ref{fig3} shows a spectrum (an effective amplitude) of
gravitational waves 
for $D=50$\,Mpc for M07 models.  Here, the Fourier spectrum $h(f)$ is
first derived from~\cite{yamamoto08}
\beqn 
h(f)=- {}_{-2}Y_{20}\int_{-t_i}^{t_f} dt {2\Psi_{20}(t) \over (2\pi f)^2}
\exp(-2\pi i f t), 
\eeqn
where $t_i=t_f \approx 10$--$30$\,ms for the choice of time in which
the maximum of $\Psi_{20}$ is reached at $t=0$.  Here, $t_i$ and $t_f$
are varied depending on the characteristic wavelength of gravitational
waves.  Since $_{-2}Y_{20}$ is proportional to $\sin^2\theta$, the
average of $h(f)$ with respect to $\theta$ is $h_{\rm
  ave}(f)=2h(f)_{\theta=\pi/2}/3$.  Then, we define the effective
amplitude by $h_{\rm eff,ave}:=f|h_{\rm ave}(f)|$~\cite{note}.  In
Fig.~\ref{fig3}, we, in addition, multiply 2 because the
signal-to-noise ratio (SNR) for a spectrum of gravitational waves, $g(f)$,
is written as
\beq
{\rm SNR}=\int_0^{\infty} df {(2|g(f)|)^2 \over S_n(f)} 
\eeq
where $S_n(f)$ is the one-sided noise spectrum density of a
gravitational-wave detector (i.e., for each frequency $f$,
$2|fh(f)|/\sqrt{S_n(f)f}$ approximately denotes the SNR).
The black dot-dot and grey dot-dot-dot curves show the design
sensitivities of advanced LIGO (the ``Zero Detuning High Power''
configuration~\cite{ligonoise}) and Einstein Telescope of the type
B~\cite{etnoise}.  Here, we plot a dimensionless quantity $(S_n
f)^{1/2}$.  We note that a low frequency part of $h(f)$ depends on the
choice of $t_i$ and $t_f$, and hence, we do not plot the unreliable
part of $h_{\rm eff,ave}$ in Fig.~\ref{fig3}.

Figure~\ref{fig3} shows that the peak amplitude of $h_{\rm eff,ave}$
is proportional approximately to the initial mass of the formed BH. 
On the other hand, the highest frequency for the peak amplitude
is inversely proportional to the BH mass. Broadly speaking,
for the initial BH mass, $M_i$, the maximum value of $2h_{\rm
  eff,ave}$ is written as $\sim 10^{-22}(M_i/30M_\odot)$ and the
corresponding highest frequency is $\sim 400(M_i/30M_\odot)^{-1}$\,Hz.
As indicated from the right panel of Fig.~\ref{fig2}, these values
depend very weakly on the angular momentum of the CO cores for 
$\chi \agt 0.5$. 

For the curves shown in Fig.~\ref{fig3}, the SNR for the most
optimistic alignment of a detector with respect to the direction of
the source (see~\cite{note} for a remark) is calculated for
$D=50$\,Mpc. For the sensitivity of advanced LIGO, it becomes $\approx
11$ for the model with no nuclear reaction and no neutrino cooling,
$\approx 3.1$ for the models with the nuclear reaction and $\rho_0=0$
and $10^{10}\,{\rm g/cm^3}$, and $\approx 1.7$ with nuclear reaction
and $\rho_0=10^{11}\,{\rm g/cm^3}$. For the sensitivity of Einstein
Telescope of the type B, on the other hand, each value is enhanced as
164, 47, 46, and 24, respectively. Obviously, for a higher value of
$M_i$, the SNR is higher. However, the most realistic model is those
with the nuclear reaction and $\rho_{0}=10^{11}\,{\rm g/cm^3}$ (for
which SNR$=1.7$).  This implies that for the second generation
detectors (advanced LIGO, advanced VIRGO, and KAGRA), it would be
difficult to detect gravitational waves emitted by the collapse of a
VMS of initial mass $\sim 300M_\odot$ unless the collapse occurs for
$D \alt 10$\,Mpc. However, with the third-generation detectors such as
Einstein Telescope, these gravitational waves could be detected with
SNR $\agt 10$ for events with $D \alt 100$\,Mpc.


The right panel of Fig.~\ref{fig3} shows the spectrum for models M01,
M03, M05, M07, and M08. This shows that for $\chi \agt 0.5$, the peak
amplitude depends weakly on the value of $\chi$, but for the small
values of $\chi$, it decreases steeply, and in addition, the amplitude
becomes steeply small with the decrease of $f$. Thus, to get a high SNR,
a moderately large value of $\chi \agt 0.5$ is necessary.

\section{Summary and discussion}

By new numerical-relativity simulations, we derived gravitational
waveforms from a rotating VMS core collapsing to a BH and found that
they are characterized by a ring-down oscillation of the formed BH in
its early formation phase. For a plausible setting of the nuclear
reaction and neutrino cooling, the initial BH mass is $\approx
20M_\odot$ for $M_{\rm CO} \approx 150M_\odot$. For the moderately
rapidly rotating case, gravitational waves have a broad peak in the
spectrum for $f \approx 300$--$600$\,Hz with $h_{\rm eff,ave}\sim
10^{-22}$ for $D=50$\,Mpc, for which the SNR for the designed
sensitivities of advanced LIGO and Einstein Telescope of the type B
would be $\alt 2$ and 20, respectively. Thus, it would be difficult to
detect such gravitational waves by the second-generation detectors,
but they will be one of the targets for the third-generation
detectors.

We should keep in mind that for very massive CO cores ($M_{\rm CO}\gg
150M_\odot$), the initial BH mass would be much higher, say
$100M_\odot$. For such a case, the peak effective amplitude of
gravitational waves would be $\sim 4 \times 10^{-22}$ and the
corresponding characteristic frequency would be $\sim 100$\,Hz (see
Fig.~\ref{fig3}).  For these gravitational waves, the SNR can be $\sim
10$ for $D=100$\,Mpc for the designed sensitivity of advanced LIGO.
Thus, for such VMS collapse, gravitational waves would have an SNR high
enough for the detection by the second-generation detectors.

As we illustrated in Ref.~\cite{Uchida18}, electromagnetic signals
could be emitted after the BH formation in the collapse of rotating
VMSs.  Detection of gravitational waves will be used for constraining
the sky location for the search of such electromagnetic signals, which
could give important information on the BH formation and subsequent
evolution of the system.


{\em Acknowledgments}: Numerical computations were performed on XC50
at CfCA of NAOJ and XC40 at YITP of Kyoto University. This work was
supported by Grant-in-Aid for Scientific Research (Grants
Nos. 16H02183, 16K17706, 16H05341, 15H00782) of Japanese MEXT/JSPS. KT
was supported by the JSPS Overseas Research Fellowships.



\begin{thebibliography}{99}
\bibitem{note} If we further take into account the antenna pattern of
  the gravitational-wave detectors with respect to the injection
  direction of gravitational waves, the root-mean square of the actual
  amplitude which the detectors measure would be by a factor
  of $1/\sqrt{5}$ smaller.
\end{thebibliography}
\end{document}